\documentclass[UTF8,a4paper]{article}
\usepackage{amsmath}
\usepackage{graphicx}
\usepackage{subfigure}
\usepackage{epstopdf}
\usepackage{geometry}
\usepackage{ulem}
\usepackage{indentfirst}
\usepackage{amsfonts,amssymb,dsfont}
\usepackage{setspace}
\usepackage{threeparttable}
\usepackage{amsmath,mathtools,amsthm}
\usepackage{array}
\usepackage{extarrows}
\usepackage{upgreek}

\makeatletter
\renewcommand*\env@matrix[1][\arraystretch]{%
  \edef\arraystretch{#1}%
  \hskip -\arraycolsep
  \let\@ifnextchar\new@ifnextchar
  \array{*\c@MaxMatrixCols c}}
\makeatother
\begin{document}


\title{\bf Josephson effect in silicene-based SNS Josephson junction: Andreev reflection and free energy}
\author{Chen-Huan Wu
\thanks{chenhuanwu1@gmail.com}
\\Key Laboratory of Atomic $\&$ Molecular Physics and Functional Materials of Gansu Province,
\\College of Physics and Electronic Engineering, Northwest Normal University, Lanzhou 730070, China}

\maketitle
\vspace{-30pt}
\begin{abstract}
\begin{large}
We investigate the Josephson effcet in superconductor-normal-superconductor junction (SNS) base on the doped unbiased silicene
under the perpendicular electric field and off-resonance circularly polarized light.
The Andreev reflection (including the retroreflection and specular one) during the subgap transport, the free energy, and the reversal of the Josephson effect
as well as the emergence of $\phi_{0}$-junction are exploited.
The Andreev reflection is complete in the NS interface even for the clean 
interface and without the Fermi wave vector mismatch, which is opposite to the case of ferromagnet-superconductor interface.
The important role played by the dynamical polarization of the degrees of freedom to the $0-\pi$ transition 
and the generation of $\phi_{0}$-junction are mentioned in this paper.
The scattering by the charged impurity in the substrate 
affects the transport properties in the bulk as well as the valley relaxation, which
can be taken into consider by the macroscopic wave function.
In short junction limit, the approximated results about the Andreev level and free energy are also discussed.
Beside the low-energy limit of the tight-binding model, the finite-size effect need to be taken into account as long as
the spacing model is much larger than the superconducting gap.

\end{large}

\end{abstract}
\begin{large}
\section{Introduction}


In this paper, we consider a silicene-based superconductor-normal-superconductor Josephson junction with
the silicene nanoribbons lies along the $k_{x}$ (zigzag) direction in a two-terminal geometry as shown in Fig.1.
The silicene is divided into three parts, the middle part is deposited on the SiO$_{2}$ substrate,
and thus with the dielectric constant $\epsilon=2.45$ ($\epsilon_{SiO_{2}}=3.9$),
while the left- and right-side parts are deposited on the conventional superconductor electrodes
with the $s$-wave superconductors realized by the superconducting proximity effect.
The perpendicular electric field and off-resonance circularly polarized light are applied on the middle normal region.
We set a finite chemical potential $\mu_{n}$ in the middle region by slightly doping,
while at the superconductive regions, the chemical potential $\mu_{s}$ required by the high carriers density 
is much larger than the Fermi wave-vector ${\bf k}_{F}$ 
and the Dirac-mass $m_{D}^{\eta\sigma_{z}\tau_{z}}$,
which has ${\bf k}_{F}=\sqrt{\mu_{s}^{2}-(m_{D}^{\eta\sigma_{z}\tau_{z}}-U)^{2}}/\hbar v_{F}$ with
$U$ the electrostatic potential induced by the doping or gate voltage in the superconducting region,
which breaks the electron-hole symmetry,
and lifts the zero-model (between the lowest conduction band and the highest valence band)
above the Fermi level (imaging the zero Dirac-mass here)\cite{Rainis D}.
The $U$ has been experimentally proved to be valid in controlling the phase shift of the $\phi_{0}$-junction
as well as the $0-\pi$ transition\cite{Szombati D B} like the bias voltage.
Note that the Dirac-mass here is related to the band gap by $\Delta=2m_{D}^{\eta\sigma_{z}\tau_{z}}$.
Thus we can know that $\mu_{s}\gtrsim \sqrt{(m_{D}^{\eta\sigma_{z}\tau_{z}}-U)^{2}}$ and the incident angle is larger than the transmission angle
due to the relation ${\bf k}_{F}{\rm sin}\theta_{n}=\mu_{s}{\rm sin}\theta_{s}$\cite{Linder J}
where $\theta_{n}$ is the incident angle from the normal region and $\theta_{s}$ is the transmission angle in the superconducting region.
Furthermore, we can estimate the transmission angle as $\theta_{s}\approx 0$ to obtain the zero scattering angle and the smooth propagation.

The tight-binding Hamiltonian of the monolayer silicene reads
\begin{equation} 
\begin{aligned}
H=&\hbar v_{F}(\eta\tau_{x}k_{x}+\tau_{y}k_{y})+\eta\lambda_{{\rm SOC}}\tau_{z}\sigma_{z}+a\lambda_{R_{2}}\eta\tau_{z}(k_{y}\sigma_{x}-k_{x}\sigma_{y})\\
&-\frac{\overline{\Delta}}{2}E_{\perp}\tau_{z}+\frac{\lambda_{R_{1}}}{2}(\eta\sigma_{y}\tau_{x}-\sigma_{x}\tau_{y})+M_{s}s_{z}+M_{c},
\end{aligned}
\end{equation}
where $E_{\perp}$ is the perpendicularly applied electric field, 
$a=3.86$ is the lattice constant,
$\overline{\Delta}$ is the buckled distance in $z$-diraction between the upper sublattice and lower sublattice,
$\sigma_{z}$ and $\tau_{z}$ are the spin and sublattice (pseudospin) degrees of freedom, respectively.
$\eta=\pm 1$ for K and K' valley, respectively.
$M_{s}$ is the spin-dependent exchange field and $M_{c}$ is the charge-dependent exchange field.
$\lambda_{SOC}=3.9$ meV is the strength of intrinsic spin-orbit coupling (SOC) and $\lambda_{R_{2}}=0.7$ meV is the intrinsic Rashba coupling
which is a next-nearest-neightbor (NNN) hopping term and breaks the lattice inversion symmetry.
$\lambda_{R_{1}}$ is the electric field-induced nearest-neighbor (NN) Rashba coupling which has been found that linear with the applied electric field
in our previous works\cite{Wu C H1,Wu C H2,Wu C H3,Wu C H4,Wu C H5}, which as $\lambda_{R_{1}}=0.012E_{\perp}$.
For circularly polarized light with the electromagnetic vector potential has ${\bf A}(t)=A(\pm{\rm sin}\ \Omega t,{\rm cos}\ \Omega t)$,
where $\pm$ denotes the right and left polarization, respectively.
Due to the perpendicular electric field $E_{\perp}$ and the off-resonance circularly polarized light which with frequency $\Omega>1000$ THz,
the Dirac-mass and the corresponding quasienergy spectrum of the normal region are
\begin{equation} 
\begin{aligned}
m_{Dn}^{\eta\sigma_{z}\tau_{z}}=&|\eta\lambda_{{\rm SOC}}s_{z}\tau_{z}-\frac{\overline{\Delta}}{2}E_{\perp}\tau_{z}+M_{s}s_{z}+M_{c}-\eta\hbar v_{F}^{2}\frac{\mathcal{A}}{\Omega}|,\\
\varepsilon_{n}=&s\sqrt{(\sqrt{\hbar^{2}v_{F}^{2}{\bf k}^{2}+(m_{Dn}^{\eta\sigma_{z}\tau_{z}})^{2}}+s\mu_{n})^{2}},
\end{aligned}
\end{equation}
respectively, where the dimensionless intensity $\mathcal{A}=eAa/\hbar$ is in a form similar to the Bloch frequency, 
and $s=\pm 1$ is the electron/hole index, and the
subscript $e$ and $h$ denotes the electron and hole, respectively.
The off-resonance circularly polarized light results in the asymmetry band gap in two valleys (see Ref.\cite{Wu C H5})
and breaks the time-reversal symmetry in the mean time,
and thus provides two pairs of the different incident electrons that may leads to the josephson current reversal
due to the valley-polarization.
The index "$n$" in above equations is to distinct them from the Dirac-mass and quasienergy spectrum in superconducting regions,
which are 
\begin{equation} 
\begin{aligned}
m_{D}^{\eta\sigma_{z}\tau_{z}}=&|\eta\lambda_{{\rm SOC}}s_{z}\tau_{z}+M_{s}s_{z}+M_{c}|,\\
\varepsilon=&s\sqrt{(\sqrt{\hbar^{2}v_{F}^{2}{\bf k}^{2}+(m_{D}^{\eta\sigma_{z}\tau_{z}})^{2}}+s \mu_{s})^{2}+\Delta_{s}^{2}},
\end{aligned}
\end{equation}
where $\Delta_{s}$ is the superconducting gap (complex pair potential) which obeys the BCS relation 
and can be estimated as $\Delta_{s}=\Delta_{0}{\rm tanh}(1.74\sqrt{T_{c}/T-1})e^{i\phi/2}$ 
(here we only consider the right superconducting lead)\cite{Zhou X,Annunziata G}
with $\phi$ the macroscopic phase-difference between the left and right superconducting leads,
$\Delta_{0}$ the zero-temperature energy gap which estimated as 0.001 eV\cite{Rainis D} here and $T_{c}$ the superconducting critical temperature
which estimated as $5.66\times 10^{-4}$ eV.
The superconducting gap is often used to compared with the excitation gap in normal region,
thus we simply use $\Delta_{s}$ to replace the $\Delta_{s}+m_{D}^{\eta s_{z}\tau_{z}}$ in below.
Obviously, the quasienergy spectrum here is distinct from the one obtained by the Floquet technique in low-momentum limit as presented in the
Refs.\cite{López A,Wu C H5}.
Note that since the exchange field considered here ($M_{s}=M_{c}=0.0039$ eV),
the critical electric field (for the zero Dirac-mass) is no more 0.017 eV/\AA,
but $\lesssim 0.051$ eV/\AA for small light-intensity.

\section{Andreev bound state}

The Andreev reflection (AR) happen in the middle normal region (or insulating barrier)
for the bias voltage smaller than $\Delta_{s}$ and $U$ (gate voltage) which been applied across the junction\cite{Rainis D}.
We consider the quasi-normal incidence of the electrons from normal region,
i.e., with constant $k_{y}$ and non-constant $k_{x}$.
The Andreev bound state, which is common in the $d$-wave superconductor\cite{Kashiwaya S},
exist in the middle normal region unless there is a band insulator,
that can be realized by control the electric field and the light field which takes effect in the middle normal region.
For SNS Josephson junction,
in contrast to the normal reflection (NR; like the 
elastic cotunneling (ECT) process in ferromagnet (normal)/superconductor/ferromagnet (normal) (NSN) junction)
whose reflection and transmission coefficients obeys the relation 
$|r_{e}|^{2}+|t_{e}|^{2}=1$, while 
the AR process obeys $|r_{h}|^{2}+|t_{e}|^{2}=1$,
where $r_{h}$ is the reflection coefficient of the reflected hole from conduction band (retro-like) or valence band (specular),
and $t_{e}$ is the transmission  coefficient of the electron-like quasiparticle.
The ECT relys on the coherent superposition of states through the, e.g., quantum dot orbit in the parallel configuration between two electrodes.
For the AR process, here the reflection coefficient reads
\begin{equation} 
\begin{aligned}
r_{h}=\frac{e^{i(\eta\theta_{n}-\phi)}{\bf k}_{F}{\rm cos}\theta_{n}}{i{\bf k}_{F}{\rm sin}\beta{\rm cos}\theta_{n}+\varepsilon_{n}{\rm cos}\beta},
\end{aligned}
\end{equation}
where $\theta_{n}={\rm asin}\frac{\mu_{s}{\rm sin}\theta_{s}}{{\bf k}_{F}}$
and $s=\pm 1$ is the electron(+1)/hole(-1) index and the Fermi wave vector can be obtained by Eq.(2)
\begin{equation} 
\begin{aligned}
{\bf k}_{F}=\frac{\sqrt{(\varepsilon_{n}-s\mu_{n})^{2}+(m_{D}^{\eta s_{z}\tau_{z}})^{2}}}{\hbar v_{F}}.
\end{aligned}
\end{equation}
We can see that the reflection probability is $|r_{h}|^{2}$ and it equas to one for vertical incidence $\theta_{n}=0$
and $\beta=0$ (i.e., $\varepsilon_{n}=\Delta_{s}$).
In Fig.1, we present the Andreev reflection probability $|r_{h}|^{2}$ versus the phase difference with different superconducting gap $\Delta_{s}$
and electric field.
We can obtained that, in the case of $\varepsilon_{n}>\Delta_{s}$,
the Andreev reflection probability decrease with the increase of Dirac-mass $m_{Dn}^{\eta s_{z}\tau_{z}}$ or $\Delta_{s}$.
The factor $\beta$ here is associate with the relation between the supra-gap or subgap excitation energy and the superconducting gap
\begin{equation} 
\begin{aligned}
\beta=\left\{
\begin{array}{rcl}
-i{\rm acosh}\frac{\varepsilon_{n}}{\Delta_{s}},&\ |\varepsilon_{n}|>\Delta_{s},\\
{\rm acos}\frac{\varepsilon_{n}}{\Delta_{s}},&\ |\varepsilon_{n}|<\Delta_{s},
\end{array}
\right.
\end{aligned}
\end{equation}
for a propagating scattering waves and a evanescent scattering waves, respectively,
and we have 
\begin{equation} 
\begin{aligned}
e^{i\beta}=\frac{\varepsilon_{n}\mp\sqrt{\varepsilon_{n}^{2}-\Delta_{s}^{2}}}{\Delta_{s}},
\end{aligned}
\end{equation}
where the sign $"\mp"$ takes $"+"$ for $|\varepsilon_{n}|<\Delta_{s}$,
and takes $"-"$ for $|\varepsilon_{n}|>\Delta_{s}$.
Then the transmission coefficient can be obtained as
\begin{equation} 
\begin{aligned}
t_{e}=\sqrt{1-\frac{|{\bf k}_{F}{\rm cos}\theta|^{2}e^{2({\rm Im}\phi-{\rm Im}(\eta\phi))}}{|\varepsilon_{n}{\rm cos}\beta+i{\bf k}_{F}{\rm sin}\beta{\rm cos}\theta|^{2}}},
\end{aligned}
\end{equation}
where ${\rm Im}$ denotes the imaginary part.
Both the reflection coefficient and transmission coefficient 
implies that the AR is complete in the NS interface (as depicted in Fig.1) even when the interface is clean (without impurity)
and without the Fermi wave vector mismatch,
that's opposite to the case of ferromagnet-superconductor interface\cite{De Jong M J M}.
In N-S junction, to ensure the AR to occur, the excitation energy in the normal region must be smaller than the band gap in the superconducting regions,
which including the proximity-induced 
superconducting gap $\Delta_{s}$\cite{Linder J},
i.e., only the subgap excitation energy is allowed,
which would leads to the coherent superposition of the electron and hole excitations in small excitation energy.
In this case of $\varepsilon_{n}<\Delta_{s}$,
electrons can enter into the superconductor lead only by forming a Cooper pair
which consisted of a electron with up-spin in K valley and a electron with up-spin in K' valley,
or it can't penetrate into the superconductor lead due to the small excitation energy\cite{De Jong M J M}.
Further, when $\varepsilon_{n}>\mu_{n}+m_{D}$ the AR is specular (interband), while it's retro-like (intraband) for $\varepsilon_{n}<\mu_{n}+m_{D}$.
For the former one, the large excitation energy also results in the thermal transport between two superconducting leads
by the propagating model\cite{Beenakker C W J}
and it's related to the Fermi distribution term ${\rm tanh}(\varepsilon_{n}/2k_{B}T)$, while for the latter one,
it contributes only to the localized model.

Then we focus on the dispersion of the Andreev bound level, which is
\begin{equation} 
\begin{aligned}
\varepsilon_{A}=s\frac{\Delta_{s}}{\sqrt{2}}\sqrt{1-\frac{A(C-{\rm cos}\phi)+s_{z}\sqrt{B^{2}[A^{2}+B^{2}-(C-{\rm cos}\phi)^{2}]}}{A^{2}+B^{2}}},
\end{aligned}
\end{equation}
where we have use the definitions
\begin{equation} 
\begin{aligned}
A=&C_{1}C_{2}+\frac{(S_{1}S_{2}(\frac{f_{2}}{f_{1}}+1)(\frac{f_{4}}{f_{3}}-1))}{4\sqrt{\hbar^{2}v_{F}^{2}k_{y}^{2}f_{2}/f_{4}+1}\sqrt{\frac{-f_{4}}{f_{3}}}\sqrt{-\hbar^{2}v_{F}^{2}k_{y}^{2}f_{1}/f_{3}+1}\sqrt{\frac{f_{2}}{f_{1}}}},\\
B=&\frac{S_{1}C_{2}(\frac{f_{3}}{2f_{1}}+\frac{1}{2})}{\sqrt{-(\hbar^{2}v_{F}^{2}k_{y}^{2}f_{1})/f_{3}+1}\sqrt{f_{2}/f_{1}}}-
\frac{C_{1}S_{2}(\frac{f_{4}}{2f_{2}}-\frac{1}{2})}{\sqrt{(\hbar^{2}v_{F}^{2}k_{y}^{2}f_{2})/f_{4}+1}\sqrt{-f_{4}/f_{3}}},\\
C=&\frac{\hbar^{2}v_{F}^{2}k_{y}^{2}S_{1}S_{2}}{\sqrt{\hbar^{2}v_{F}^{2}k_{y}^{2}f_{2}/f_{4}+1}\sqrt{-f_{4}/f_{3}}\sqrt{-(\hbar^{2}v_{F}^{2}k_{y}^{2}f_{1})/f_{3}+1}\sqrt{f_{2}/f_{1}}}\\
&-[1\cdot\Theta(\varepsilon_{n}-\mu_{n}-m_{D}^{\eta s_{z}\tau_{z}})+(-1)\cdot\Theta(-\varepsilon_{n}+\mu_{n}+m_{D}^{\eta s_{z}\tau_{z}}]\\
&\times\frac{(S_{1}S_{2}(f_{2}/f_{1}-1)(f_{4}/f_{3}+1))}{4\sqrt{\hbar^{2}v_{F}^{2}k_{y}^{2}f_{2}/f_{4}+1}\sqrt{-f_{4}/f_{3}}\sqrt{-(\hbar^{2}v_{F}^{2}k_{y}^{2}f_{1})/f_{3}+1}\sqrt{f_{2}/f_{1}}},
\end{aligned}
\end{equation}
with the Heaviside step function $\Theta$
which distinguish the two kinds of AR: retroreflection and specular AR,
and thus makes this expression valid for both of these two case.
The wave vectors $f_{1}\sim f_{4}$ and parameters $C_{1},\ C_{2},\ S_{1},\ S_{2}$ are defined as
\begin{equation} 
\begin{aligned}
f_{1}=m_{Dn}^{\eta\sigma_{z}\tau_{z}}+\varepsilon_{n}+\mu_{s},\ f_{2}=m_{Dn}^{\eta\sigma_{z}\tau_{z}}+\varepsilon_{n}-\mu_{s},\\
f_{3}=\varepsilon_{n}-m_{Dn}^{\eta\sigma_{z}\tau_{z}}+\mu_{s},\ f_{4}=m_{Dn}^{\eta\sigma_{z}\tau_{z}}-\varepsilon_{n}+\mu_{s},\\
C_{1}={\rm cos}(L\sqrt{f_{1}f_{3}/\hbar^{2}v_{F}^{2}-k_{y}^{2}}),\\
C_{2}={\rm cos}(L\sqrt{-f_{2}f_{4}/\hbar^{2}v_{F}^{2}-k_{y}^{2}}),\\
S_{1}={\rm sin}(L\sqrt{f_{1}f_{3}/\hbar^{2}v_{F}^{2}-k_{y}^{2}}),\\
S_{2}={\rm sin}(L\sqrt{-f_{2}f_{4}/\hbar^{2}v_{F}^{2}-k_{y}^{2}}).
\end{aligned}
\end{equation}
The $x$-component of the wave vector for the electron channel and hole channel, $k_{xe}$ and $k_{xh}$,
are incorporated in the above wave vectors,
specially, the electron and hole wave vectors here are both complex,
which implies the inclusion of the subgap solutions with the evanescent scattering waves\cite{Rainis D},
and it has $\frac{\Delta_{s}}{2\varepsilon_{n}}2{\rm cos}\beta=1$ for $|\varepsilon_{n}|<\Delta_{s}$.
Note that we only consider the dc Josephson effcet in the thermodynamic equilibrium state
here rather than the ac Josephson effcet which with time-dependent phase difference
(e.g., by a time-dependent bias voltage).

There exist a critical angles for AR process.
When the incident angle of the scattering wave excess this critical angle, it becomes exponentially-decayed.
The critical angle for AR process is
\begin{equation} 
\begin{aligned}
\theta_{AR}={\rm asin}\frac{f_{2}(-f_{4})}{f_{1}f_{3}},
\end{aligned}
\end{equation}
when the quasienergy and chemical potential $\mu_{n}$ are much smaller than the Dirac-mass $m_{Dn}^{\eta s_{z}\tau_{z}}$,
the above critical angle can be simplified as $\theta_{AR}={\rm asin}\frac{(-f_{4})}{f_{3}}$,
which is consistent with the result of Ref.\cite{Linder J2}.

While for the critical angle of ECT in the NSN junction, it is related to the electron/hole index: $s=1$ for the charge channel 
which corresponds to the parallel configuration\cite{Linder J2},
and $s=-1$ for the hole channel which corresponds to the antiparallel configuration.
The critical angle of ECT can be written as 
\begin{equation} 
\begin{aligned}
{\rm sin}\theta_{ECT}=\frac{f_{3}\delta_{s,1}+f_{1}\delta_{s,-1}}{f_{1}}.
\end{aligned}
\end{equation}
While for the case of heavily doping in the middle normal region, that's $\mu_{n}\gg \varepsilon_{n}$,
then the AR is dominated by the retro one.
In this case,
when the chemical potential is comparable with the Dirac-mass,
the AR process is suppressed by the destructive interband interference\cite{Golubov A A} among the Cooper pairs due to the large
Dirac-mass (and thus the large band gap) which brings a large dissipative effect and reduce the AR.
The minimum free energy in this SNS planar junction is $\phi$- and temperature-dependent,
and local st the 0 or $\pi$-junction.
With the defined incident angle $\theta_{s}$ for the quasiparticle injected from the one of the superconducting electrodes to the middle normal region,
the minimum free energy with AR process can be written as
\begin{equation} 
\begin{aligned}
E(\phi,T)=-k_{B}T\sum_{\eta s_{z}\tau_{z}}\int^{\pi/2}_{-\pi/2}D{\rm ln}[2{\rm cosh}(\frac{\varepsilon_{A}}{2k_{B}T})]{\rm cos}\theta_{n} d\theta_{n}.
\end{aligned}
\end{equation}

The reversal of the Josephson effect will be rised by the valley, spin, and pseudospin polarization,
which with dramatic dipole oscillation between the two components induced by the off-resonance circularly polarized light
as we have discussed\cite{Wu C H5}, and results in a nonzero center-of-mass (COM) wave vector,
just like the Josephson current realized by a ferromagnetic middle silicene 
(in superconductor/ferromagnet/superconductor ballistic Josephson junction\cite{Annunziata G}).
It's found that the oscillation of valley polarization
is related to the carrier-phonon scattering due to the photoexcitation which has a relaxation
time in picosecond range\cite{Kumar S} (since the frequency of light is setted in the terahertz range) and larger than that of the electron-eletron scattering.
On the other hand, since the silicene in normal region is been deposited on a SiO$_{2}$ substrate,
the relaxation of the valley-polarization is also associated with the screened scattering by the charged impurities within the substrate.
It's found that for a certain concentration of the impurity and the vertical distance to the middle silicene layer $d$,
the relaxation time can be obtained, which is in a dimensionless form
\begin{equation} 
\begin{aligned}
\tau=\left[\frac{n_{{\rm imp}}\varepsilon_{n}}{\hbar^{3}v_{F}^{2}}
\left(\frac{\frac{2\pi e^{2}}{\epsilon_{0}\epsilon {\bf q}}}{1+N_{f}\frac{2\pi e^{2}}{\epsilon_{0}\epsilon {\bf q}}\Pi({\bf q},\omega)}\right)^{2}\right],
\end{aligned}
\end{equation}
where $\epsilon_{0}$ and $\epsilon$ are the vacuum dielectric constant and background dielectric constant, respectively.
$N_{f}$ is the number of the degenerate which can be treated as $N_{f}=g_{s}g_{v}=4$ here.
${\bf q}$ is the scattering wave vector and $\Pi({\bf q},\omega)$ is the dynamical polarization within the random-phase-approximation (RPA)
which contains the screening effect
by the high energy state with large charge density of state (DOS) $D=W|f_{1}|/(\pi\hbar v_{F})$
where $W$ is the width of the silicene ribbron.
For the case that the screening is ignored, the relaxation time has a simply relation wih the distance to the impurity,
$\tau\propto e^{d/\ell}$\cite{Boross P},
where $\ell=0.47$ \AA for silicene.
For ballistic Josephson junction, the diffusive effect is not considered,
however, due to the existence of the screened (or unscreened) charged impurities in the substrate,
the diffusive effect (e.g., to the conductivity or the valley-polarization) need to be taken into accout.

Note that we don't consider the edge states in the wave vector space,
however, if the middle normal region is replaced by a ferromagnetic one, e.g.,
by applying the out-of-plane ferromagnetic exchange field on both the upper edge and the lower edge of middle region,
the junction will always be the 0($\pi$)-junction (unless the length of the upper edge and lower edge is unequal)
when without applying the electric field or off-resonance light.
That's because the out-of-plane ferromagnetic exchange field won't gap out the edge models (when without the Rashba-coupling)
while the in-plane ferromagnetic exchange field\cite{Rachel S,An X T} which can easily gap out the gapless edge models
even in the absence of Rashba-coupling.
In such case, the Curie temperature is required to much larger than the superconducting critical temperature
to prevent the dramastic variety of the dipole polarization.
If we apply an in-plane ferromagnetic exchange field,
the gapless edge model will be gapped out and the time revesal invariance will be broken,
and thus the edge states-supported Josephson effect disappear.
While for the out-of-plane antiferromagnetic exchange field, which may leads to the single valley characteristic,
both the edge and bulk state support the Josephson effect in the normal region\cite{Zhou X2}.
In this case, the Andreev level of the upper edge and lower edge can be easily obtained just by inversing the related degrees of freedom,
due to the chiral character-induced opposite phase between the upper edge and lower edge,
e.g., the helical edge model in quantum spin Hall phase which with up- and down-spin carriers flow toward opposite directions
in each edge;
while for the chiral edge model in the quantum anomalous Hall phase, the edge models with up- and down-spin carriers flow toward the same direction 
in each edge.
Further, for the a ferromagnetic middle region, the spin-rotation symmetry no more exist,
and thus leads to the anomalous conductance\cite{De Jong M J M}.


In short junction case ($L\le 50$ \AA), the Andreev level can be approximatively obtained as
\begin{equation} 
\begin{aligned}
\varepsilon_{AR}^{*}=\Delta_{s}{\rm cos}\left[\frac{1}{2}\left(\phi-L(\frac{f_{3}}{\hbar v_{F}}+\frac{f_{2}}{\hbar v_{F}})\right)\right],
\end{aligned}
\end{equation}
which can be further simplified as $\varepsilon_{AR}^{*}=\Delta_{s}{\rm cos}(\phi/2)$ in the case that the $\varepsilon_{n}$ is small
and that's consistent with the result of Refs.\cite{Kulik I O,Fu L},
where the transmission coefficient equals 1 here due to the partial Klein tunneling of Cooper pairs
which happen at zero Dirac-mass\cite{Missault N} (while the AR doesn't requires the zero Dirac-mass).

\section{Results and discussion}

Due to the presence of the degrees of freedom $\eta,\ s_{z},\ \tau_{z}$ 
and the electron/hole index $s$ in the above expression of the Andreev bound level,
it should has $2^{4}=16$ levels under a certain electric field and light field,
however, there may be less levels due to the degenerative case.
The Josephson current is a nonlocal supercurrent carried by the cooper pairs 
in the superconducting leads (while it's by quasiparticles in the middle normal region),
it can be found
in such SNS junction,
and its slope is affected by the dynamical polarizations of the degrees of freedom as mentioned above\cite{Wu C H5},
for the case of, e.g., $\sqrt{\varepsilon_{n}-m_{Dn}^{+++}}\neq\sqrt{\varepsilon_{n}-m_{Dn}^{---}}$.
For simplicity, we only present the result of one of the levels, where all the degrees of freedom are takes 1.
For a Cooper pair penetrated into the bulk state,
the macroscopic wave function is affected by the scattering process by the charged impurity,
and thus has $\Psi(x)\propto e^{-x/\xi}e^{d/\ell}e^{i\phi/2}$,
where $\xi$ is the superconducting coherance length and $x<L$ is the mean free path in the bulk region.
The Josephson effect here only consider the low-temperature condition ($\sim T_{c}$),
which is dominated by the elastic scattering,
while at higher temperature, the
rised inelastic scattering may leads to the switching of the Fermi parity\cite{Fu L},
and the
frequency-dependent noise which is induced by the current-current correlation in nonequilibrium Josephson 
setup leads to the $4\pi$ period of the Josephson current due to the existence of Majorana
bound states. 
That also implies that the dissipation (related to the Dirac-mass) plays a important role during the quasiparticle transportation.

For simulation, we set the parameters as follows:
The frequency of the off-resonance light is setted larger than 3000 THz which is much higher than the critical value.
The critical value here is $3t=4.8$ eV$=1200$ THz for silicene.
In the case that frequency is $3000$ THz, 
the dimensionless intensity is $\mathcal{A}=0.3$ which is much larger than the SOC parameter $\lambda_{SOC}$.
The length of the normal region $L$ is much shorter than the superconducting coherence length $L\ll\frac{\hbar v_{F}}{\Delta_{0}}$,
i.e., $L\ll 5350$ \AA, where we estimate $\hbar v_{F}=\frac{\sqrt{3}}{2}at=5.35$ here.
The $y$-component of the momentum as well as the off-resonance circularly polarized light (which we only use the right-polarization from now on)
is conserved due to the translational invariance, thus it's a good quantum number in the computation,
and we set $k_{y}=2$ meV.
${\bf k}$ is in an order of $0.01$ eV here.
While the detail setting of the parameters are labeled in the each following plot.

Fig.3 depicts the dispersion of the Andreev bound level for the AR process in 0 state
(we don't consider the breaking of inversion symmetry by the Rashba-coupling here since it's very small).
We can easily see than the period of Andreev level is $4\pi$, 
which obeys the generate relation $\varepsilon_{AR}\propto {\rm cos}(\phi/2)$\cite{Zhou X2},
while the period of $\varepsilon_{AR}/\Delta_{s}$ is $2\pi$
as shown in Fig.4, which is consistent with the results of Ref.\cite{Annunziata G}.
We detect the effects of the electric field, off-resonance circularly polarized light, chemical potential, and length of the normal region
to the Andreev bound level numerically.
We found that,
in condition: $L=2000$ \AA ,
the slope of the Andreev level is always negative in the range [$0,\pi)$ 
and positive in the range $[\pi,2\pi]$, i.e., in the 0$(\pi)$-junction
(which depends on the product between the sign and slope of the Andreev level).
From Fig.3(b), we can see that the amplitude of Andreev level exhibits non-monotonous change with the increase of length $L$.
In this case, when we increase the light-intensity parameter $\mathcal{A}$,
the slope of the Andreev level changes,
and we found the $\mathcal{A}$ should be $<0.7$ in the 
condition: $E_{\perp}=0.034$ eV, $\mu_{n}=2$ eV, $L=2000$ \AA,
for a available Andreev level (or the slope is nearly vanish for $\mathcal{A}\ge 0.7$).
The anomalous Josephson effect as well as the $\phi_{0}$-junction can be found in Fig.3(b).
From the blue dash-dot line ($L=1500$ \AA) and green line ($L=1200$ \AA) in Fig.3(b),
we can see that the sign reversal happen even in the range $[0,\pi)$,
this phenomenon emerges also in yellow line in Fig.3(a) which with $E_{\perp}=2$ eV.
While for other levels, the sign change can happen only at $\phi=\pi$.

In Fig.4(a), we present the Andreev level $\varepsilon_{AR}/\Delta_{s}$ whose period is $2\pi$\cite{Zhou X}.
The slope of the blue line,
which corresponds to the condition: $E_{\perp}=0.01$ eV, $\mu_{n}=0.03$ eV, $\mathcal{A}=0.03$, $L=2000$ \AA,
is opposite to the green line
which corresponds to the condition: $E_{\perp}=0.26$ eV, $\mu_{n}=0.03$ eV, $\mathcal{A}=0.02$, $L=2000$ \AA.
That implies the occurence of $0-\pi$ transition which with the reversed current compared to the usual $0$-junction.
Thus we can obtain that the change of electric field and the intensity of light is valid for the generation of $0-\pi$ transition 
because of the sublattice degree of freedom and the valley degree of freedom consider here, respectively.
Thus if the subalttice degree of freedom of electric field is not considered, the variety of electric field won't support the $0-\pi$ transition
or the transition to the $\phi_{0}$-junction, which is consistent with the result in Ref.\cite{Zhou X2,Zhou X}.

In the absence of the electric field, off-resonance light and the antiferromagnetic exchange field, it is always be the 0-junction
due to the presence of time-reversal invariance (and chiral symmetry),
and the Josephson supercurrent vanishes at $\phi=n\pi$ ($n$ is integer) in this case.
However, the broken symmetry or chiral leads to the nonzero supercurrent at $\phi=n\pi$ and the $\phi_{0}$-junction
which can also be implemented by a external magnetic field in nanowire quantum dots-based junction (SQUID)\cite{Szombati D B}.

Fig.5 shows the free energy versus phase difference, we can see that change of electric field or the 
length of normal region are also valid for the generation of $\phi_{0}$ transition,
which with a phase shift $\phi_{0}\in(0,\pi)$ (or $\in (\pi,2\pi)$) for the minimum free energy.
The period of free energy is $2\pi$, which is consistent with the theory\cite{Radovi? Z} 
and experimental\cite{Baselmans J J A} results, and it's
similar to the Josephson current whose period is $2\pi$ in thermodynamic equilibrium state\cite{Kwon H J},
and becomes $4\pi$ in nonequilibirum state due to the majorana bounding\cite{Wu C H6}.
Note that the free energy here is related to the width $W$ of the nanoribbon
due to the finite-size effect.
We find that the 0-$\pi$ transition and the $\phi_{0}$-junction can be realized
by changing the length of the normal region or the electric field.
The approximated Andreev level for short length $L$ is shown in Fig.6 according to Eq.(16),
where the $0-\pi$ transition (see the black line and gray line) and the $\phi_{0}$-junction can be implemented
by changing the electric field, $\mathcal{A}$, and the length $L$.
Phenomenologically, through Fig.6, the Eq.(16) can be rewritten as 
$\varepsilon_{AR}^{*}=\Delta_{s}{\rm cos}\left[\frac{1}{2}(\phi+\phi_{0})\right]$ with $\phi_{0}\in[0,2\pi)$,
the $\phi_{0}$ here is controlled by the variables presented in the plot,
and this expression is similar to that of anomalous switching current\cite{Szombati D B}.
The usual current-phase relation can be deduced from the derivative of $\varepsilon_{AR}^{*}$
as $J=-\frac{2e}{\hbar}{\rm sin}\left[\frac{1}{2}(\phi+\phi_{0})\right]$ in low-temperature limit.
Note that these results (including the anomalous Josephson effects) are all base on the first-order perturbation theory
with the perturbation term $V$ which couples the left and right superconducting leads.

The free energy of the approximated Andreev level in Fig.6 is presented in Fig.7,
where we show the free energy in one period, 2$\pi$.
It's obvious that the free energy exhibits the characteristic of the $\phi_{0}$-junction
with the minimum free energy 
(i.e., the maximum value of the -$E(\phi,T)$ in the plot) 
change their positions under different parameters.
By comparing the black, purple, yellow lines in Fig.7(a), we can obviously see the phase shift induced by the variety of length of normal region.
While for the temperature near (but larger than) the critical one $T_{c}$,
as shown in Fig.7(b),
it doesn't affects the existence of 0($\pi$)-junction and the time reversal invariance,
and thus the minimum free energy is always localed at $\phi=3.5$.
For temperature lower than the critical one $T_{c}$ as shown in Fig.6(c),
the minimum free energy localed at $\phi=5.24$ for temperature lower than 0.6$T_{c}$,
while it localed at $\phi=5.48$ for temperature larger or equals 0.6$T_{c}$ as shown in Fig.6(b).
In contrast to the Rashba-coupling-induced helical $p$-wave spin-triplet superconductor,
the Josephson current of conventional $s$-wave superconductor saturates at low-temperature (e.g., at $T<0.6T_{c}$)\cite{Asano Y},
and the free energy doesn't change gradually with the variational temperature.

\section{Conclusion}

In the interface between the normal nanoribbon (strip) region and the superconducting region,
the AR is related to the width $W$ (the number of edge states along the armchair direction) of the ribbon:
For $W$ is even, the AR supressed by the opposite pseudospin degree of freedom between first sublattice and last subalttice in the $k_{y}$ direction,
while for $W$ is odd, the AR is allowed.
In this strip scenario,
the model spacing between bands above the Dirac level depends on $W$ and the NN hopping when consider the finite-size effect,
the model spacing is $\delta\Delta=3\pi t/(3W-2)$ for $W$ is even and $\delta\Delta=\pi t/(W-1)$ for $W$ is odd.
The band structure of the zigzag silicene nanoribbon in strip geometry is shown in Fig.8,
where the model spacing $\delta\Delta$ is indicated in the inset.
The finite-size effect here is important in large energy range,
but it can be ignored for low-energy limit as the tight-binding model depicted in Sec.1.

In this paper we investigate the Josephson effcet in superconductor-normal-superconductor junction base on the doped silicene
with a dc Josephson current in zero voltage.
We found that the dynamical polarizations of the degrees of freedom mentioned above can induce
the $0-\pi$ transition and the emergence of the $\phi_{0}$-junction.
For example, the change of the electric field or off-resonance circularly polarized light may induce the $0-\pi$ transition by the pseudospin degree of freedom
or valley degree of freedom, respectively;
the interaction between antiferromagnetic exchange field and the SOC may induce the $0-\pi$ transition by the valley polarization\cite{Zhou X2};
the interaction between the internal exchange field and Josephson 
superconducting current may induce the $\phi_{0}$-junction when the normal region is noncentrosymmetric\cite{Buzdin A}.

\end{large}
\renewcommand\refname{References}

\clearpage

Fig.1
\begin{figure}[!ht]
   \centering
 \centering
   \begin{center}
     \includegraphics*[width=0.5\linewidth]{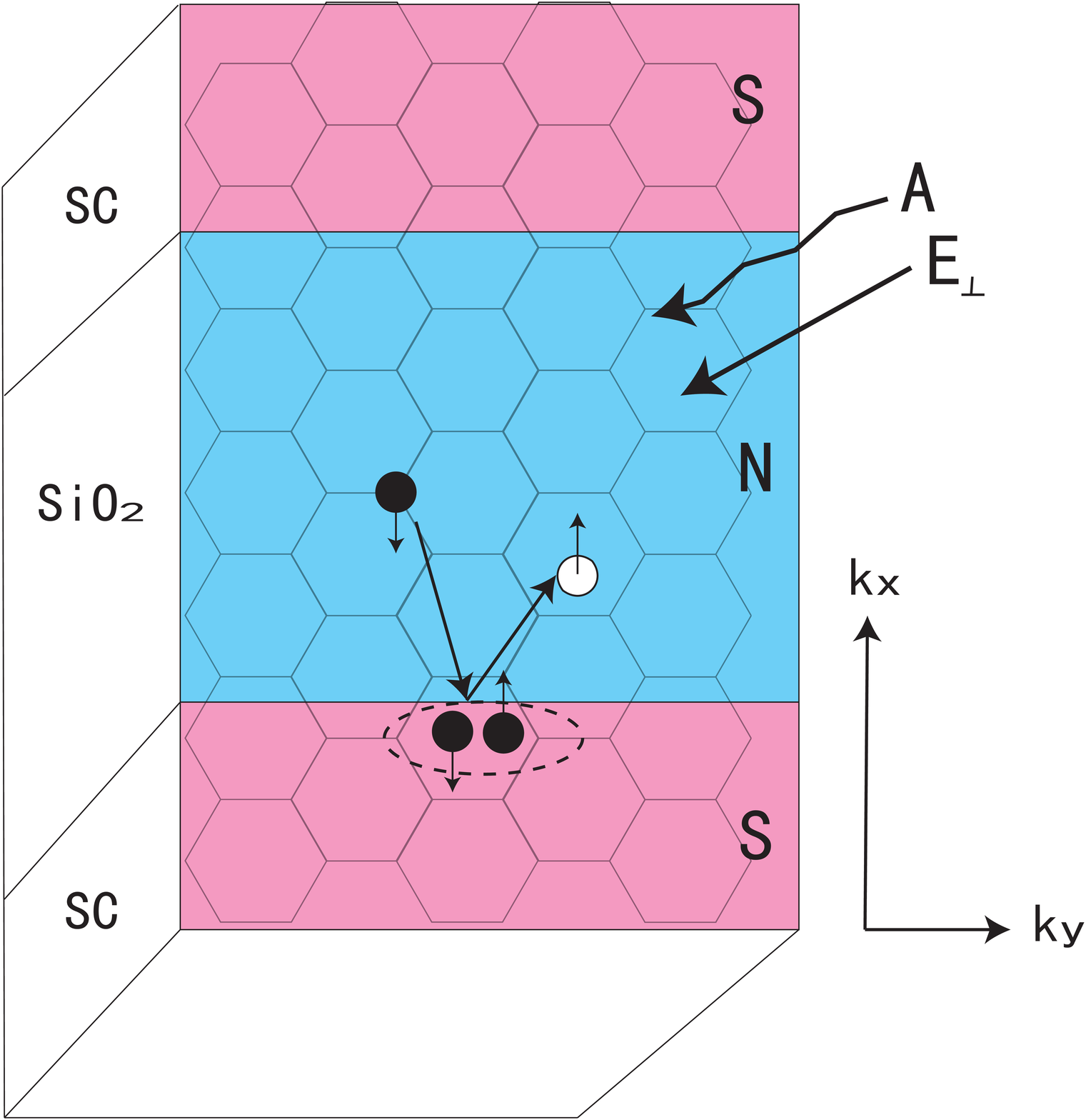}
\caption{(Color online) Schematic of the
silicene-based superconductor-normal-superconductor Josephson junction (SNS) with
the silicene nanoribbons lies along the $k_{x}$ (zigzag) direction.
The middle normal region is deposited on the SiO$_{2}$ substrate,
while the left- and right-side parts are deposited on the conventional superconducting electrodes (SC)
with the $s$-wave superconductors realized by the superconducting proximity effect.
The electric field (straight arrow) and off-resonance circularly polarized light (broken arrow) are applied perpendicularly on the middle normal region.
The process of the Andreev reflection is shown,
where the solid black circle stands electron and the hollow circle stands hole and with the short arrows stands the direction od spin.
We can see that an electron with down-spin from the conduction band of the normal region incident into the NS interface and reflected as a hole with up-spin
from the valence band (specular Andreev reflection),
and forms a Cooper pair (within dash-line circle) together with an electron with up-spin generated by the reflected hole.
}
   \end{center}
\end{figure}
\clearpage

\clearpage
Fig.2
\begin{figure}[!ht]
   \centering
 \centering
   \begin{center}
     \includegraphics*[width=0.5\linewidth]{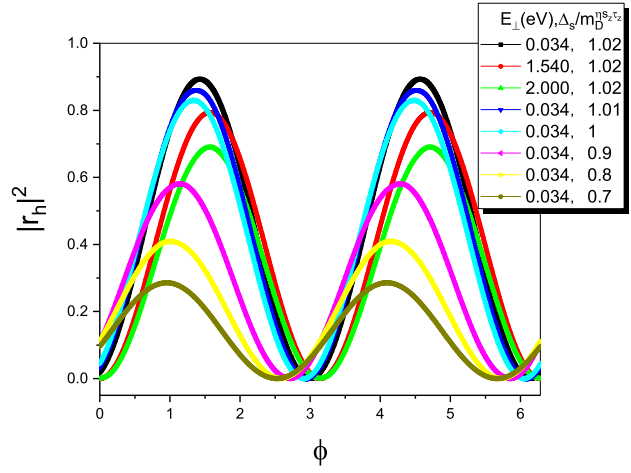}
\caption{(Color online) Andreev reflection probability $|r_{h}|^{2}$ versus the phase difference with different superconducting gap $\Delta_{s}$
and electric field.
The setting of the parameters are: $\mathcal{A}=0.3$, $\mu_{n}=0.2$ eV, $L=50$ \AA.
The superconducting gap $\Delta_{s}$ here is setted as always smaller than $\varepsilon_{n}$ thus $\beta=-i{\rm acosh}\varepsilon_{n}/\Delta_{s}$.
The Dirac-mass in the normal region $m_{Dn}^{\eta s_{z}\tau_{z}}$ are
0.2108 eV, 0.5572 eV, 0.663 eV for $E_{\perp}=0.034$ eV, $E_{\perp}=1.54$ eV, $E_{\perp}=2$ eV, respectively.}
   \end{center}
\end{figure}
\clearpage
Fig.3
\begin{figure}[!ht]
   \centering
\subfigure{
\begin{minipage}[t]{0.4\textwidth}
\includegraphics[width=1\linewidth]{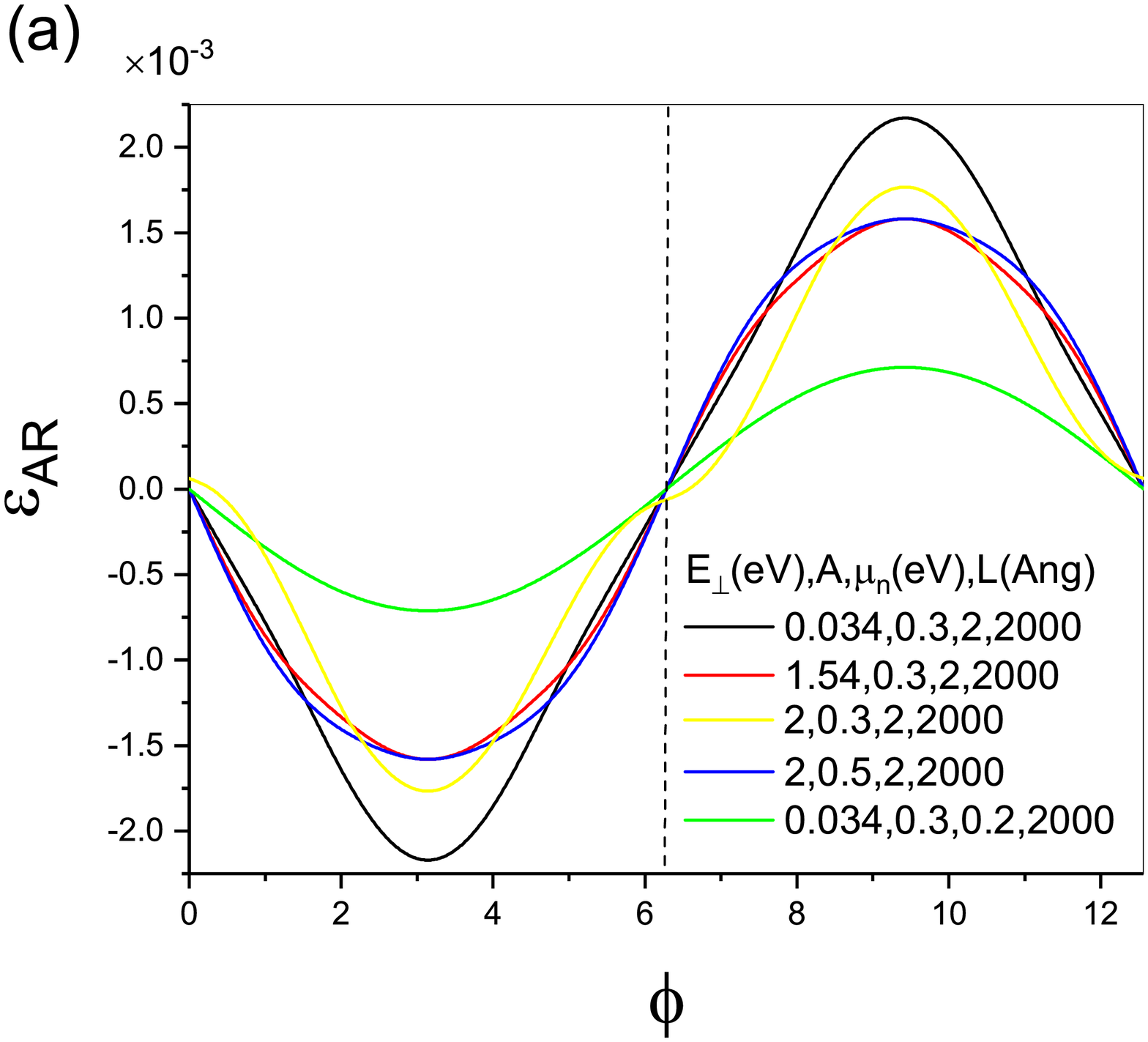}
\label{fig:side:b}
\end{minipage}
}\\
\subfigure{
\begin{minipage}[t]{0.4\textwidth}
\includegraphics[width=1\linewidth]{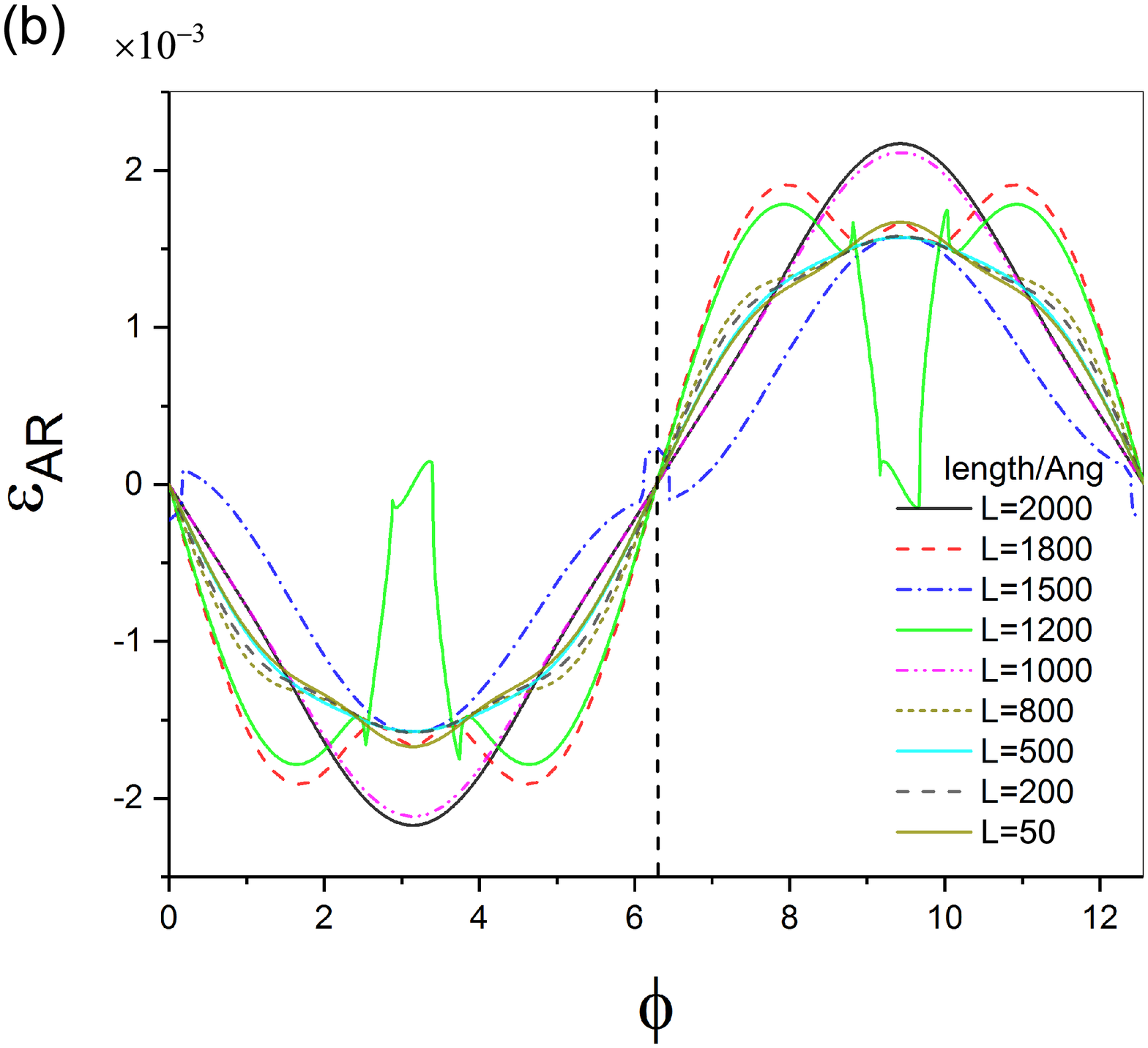}
\label{fig:side:a}
\end{minipage}
}
\caption{(Color online) Andreev bound level 
versus phase difference $\phi$.
The dash-line indicate the half-period at $\phi=2\pi$.
The spin-dependent exchange field $M_{s}$ and the charge-dependent exchange field $M_{c}$ are all setted as 0.0039 eV.
For each curve, the corresponding setting of the parameters are labeled in the plot.
}
\end{figure}
\clearpage
Fig.4
\begin{figure}[!ht]
   \centering
\subfigure{
\begin{minipage}[t]{0.4\textwidth}
\includegraphics[width=1\linewidth]{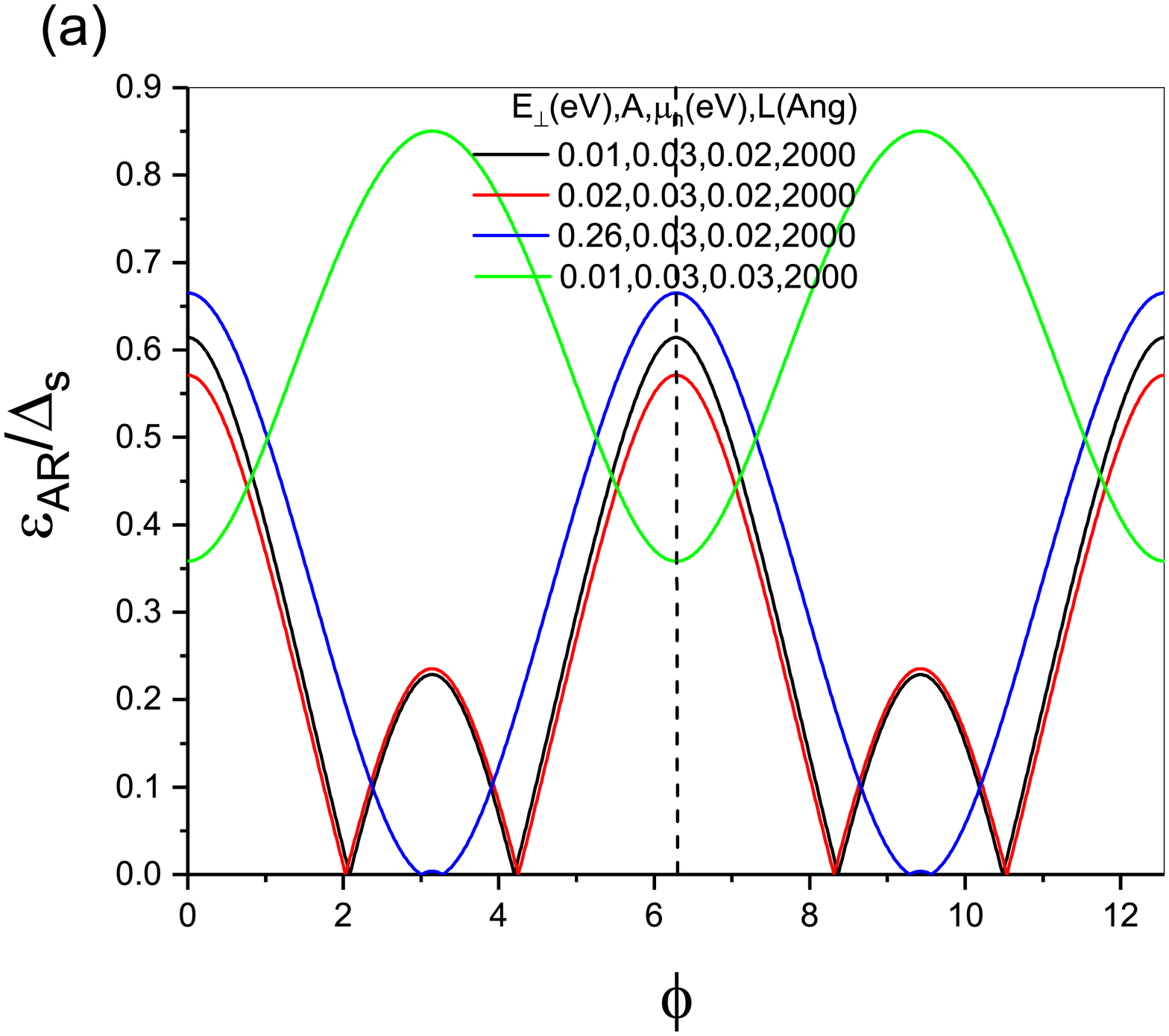}
\label{fig:side:b}
\end{minipage}
}\\
\subfigure{
\begin{minipage}[t]{0.4\textwidth}
\includegraphics[width=1\linewidth]{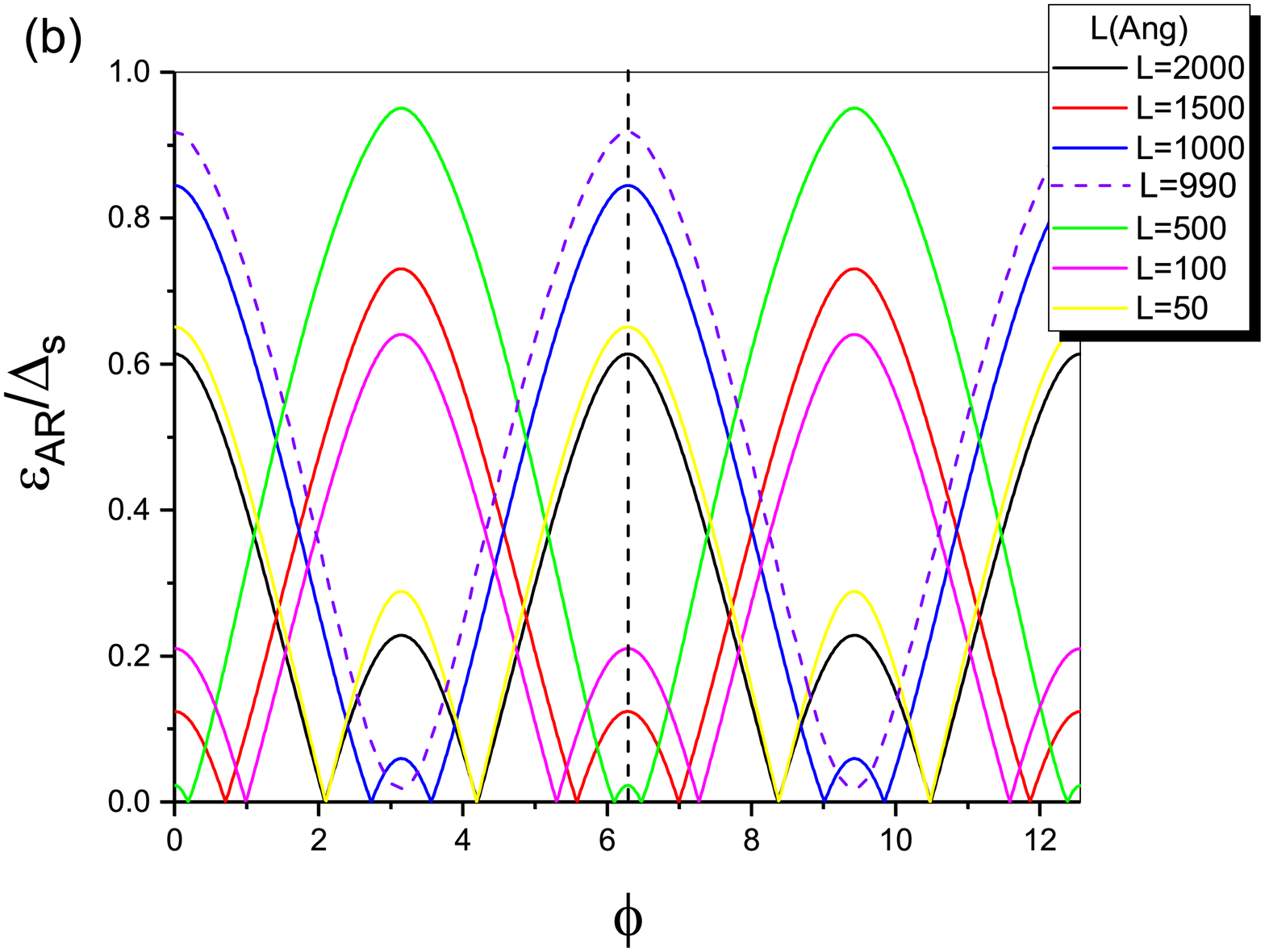}
\label{fig:side:a}
\end{minipage}
}
\caption{(Color online) $\varepsilon_{AR}/\Delta_{s}$ versus versus the phase difference $\phi$.
For each curve, the corresponding setting of the parameters are labeled in the plot.
}
\end{figure}
\clearpage
Fig.5

\begin{figure}[!ht]
   \centering
\subfigure{
\begin{minipage}[t]{0.4\textwidth}
\includegraphics[width=1\linewidth]{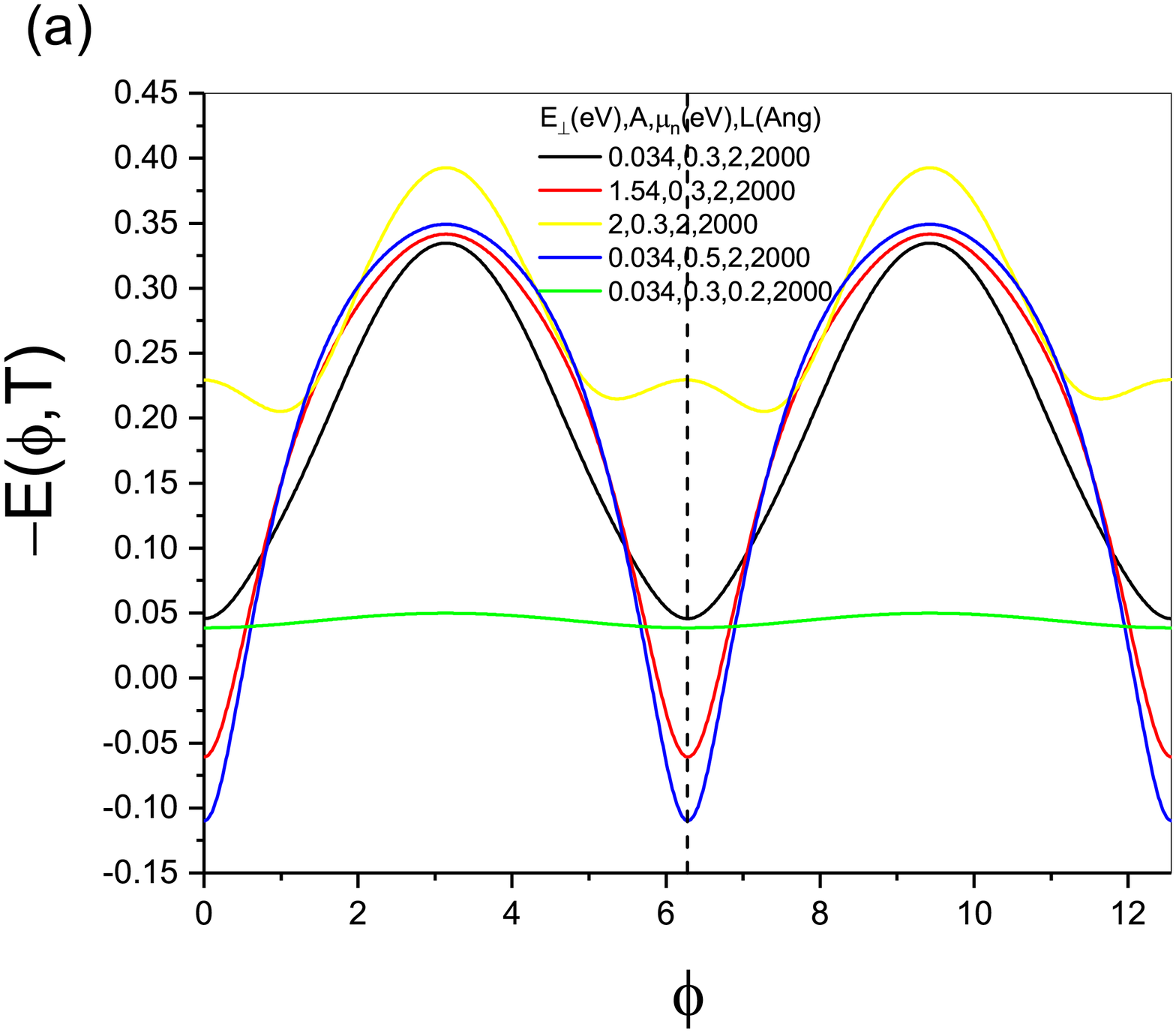}
\label{fig:side:b}
\end{minipage}
}\\
\subfigure{
\begin{minipage}[t]{0.4\textwidth}
\includegraphics[width=1\linewidth]{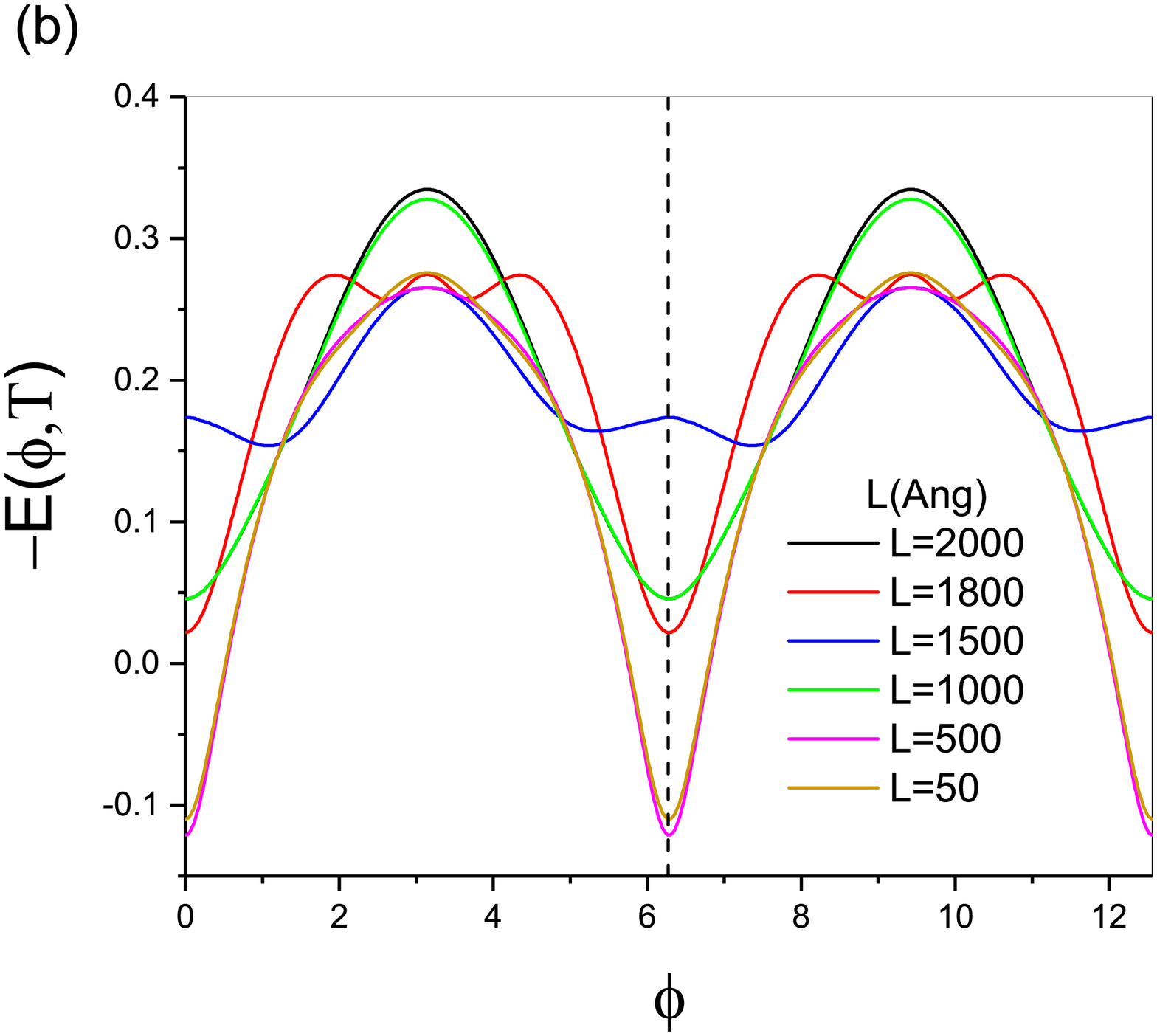}
\label{fig:side:a}
\end{minipage}
}
\caption{(Color online) Free energy versus the phase difference. The corresponding setting of the parameters are labeled in the plot.
Here we set width of ribbon as $W=126$.}
\end{figure}
\clearpage
Fig.6
\begin{figure}[!ht]
   \centering
   \centering
   \begin{center}
     \includegraphics*[width=0.5\linewidth]{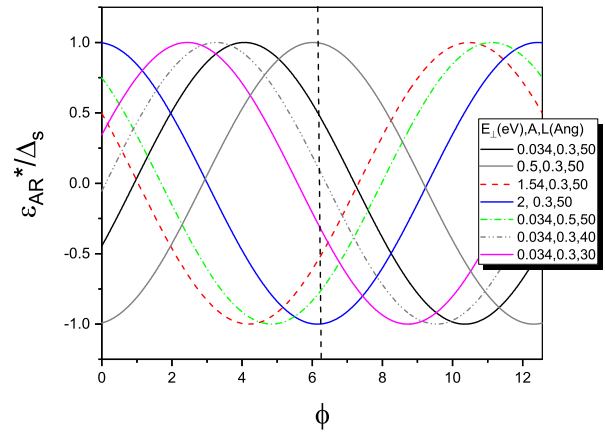}
\caption{(Color online) Andreev bound level versus phase difference according to the approximation result of Eq.(16) with short
junction length. The temperature setted here is 0.5$T_{c}$. The corresponding setting of the parameters are labeled in the plot.
}
   \end{center}
\end{figure}
\clearpage
Fig.7
\begin{figure}[!ht]
   \centering
\subfigure{
\begin{minipage}[t]{0.4\textwidth}
\includegraphics[width=1\linewidth]{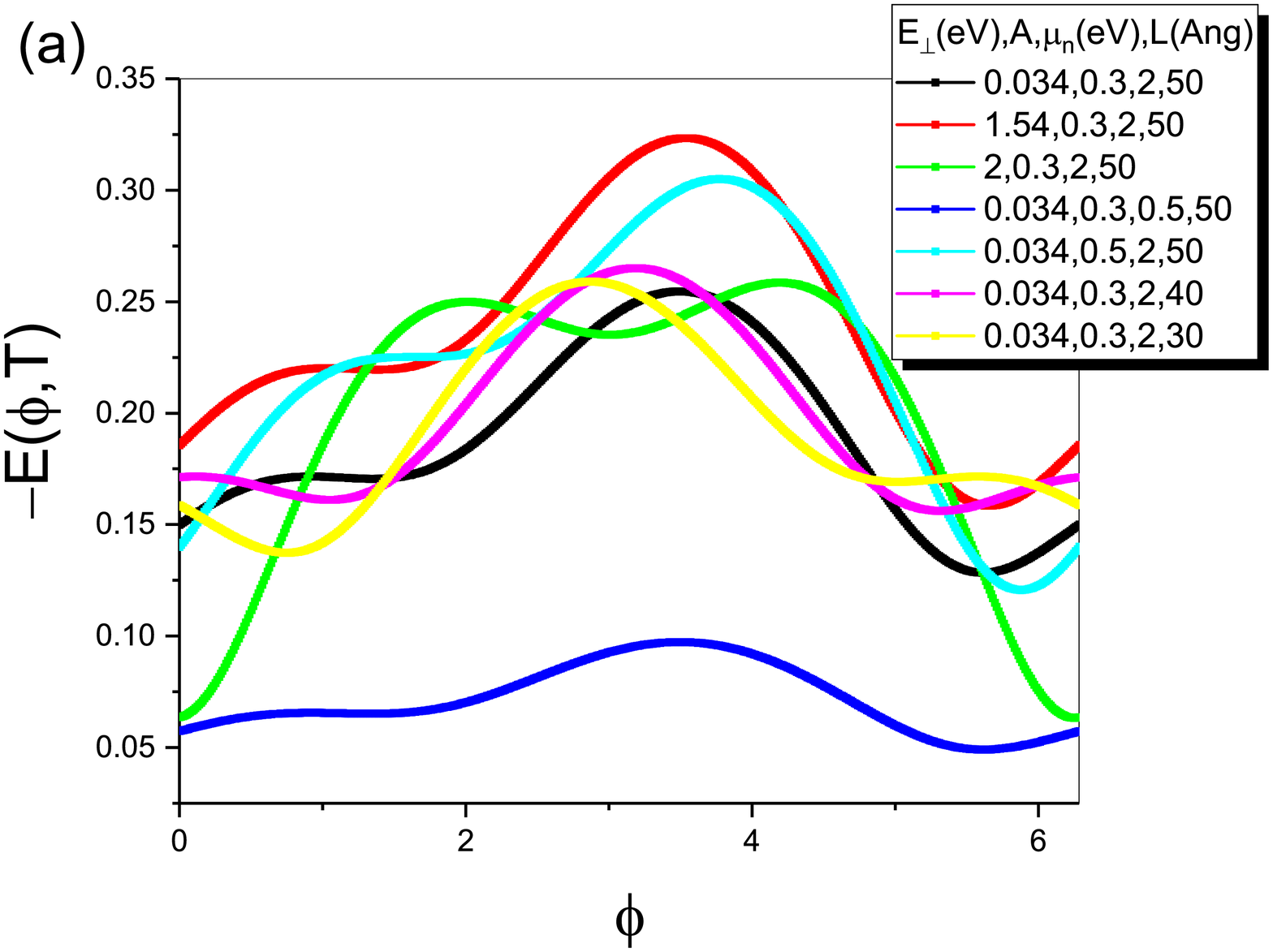}
\label{fig:side:b}
\end{minipage}
}\\
\subfigure{
\begin{minipage}[t]{0.4\textwidth}
\includegraphics[width=1\linewidth]{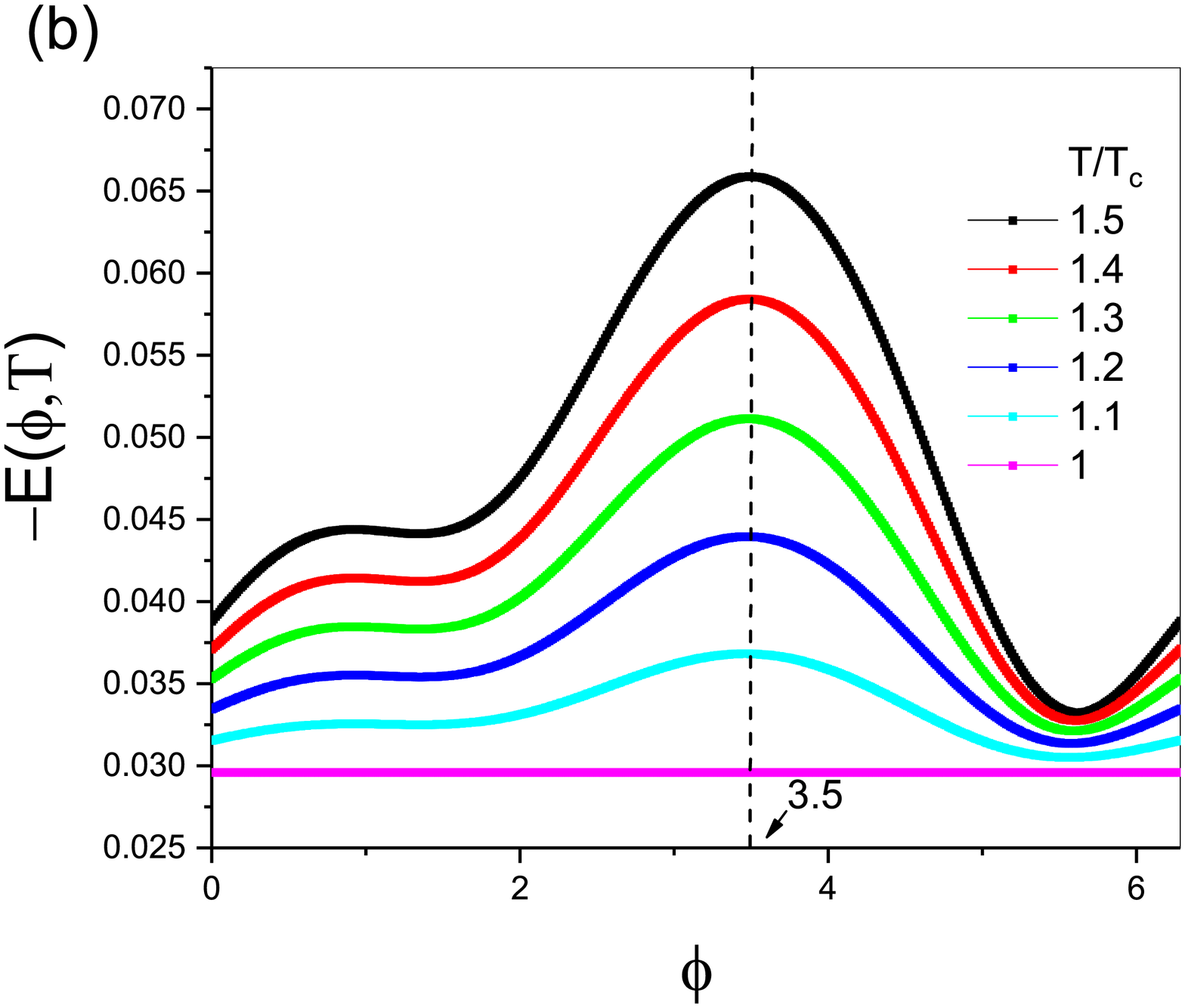}
\label{fig:side:a}
\end{minipage}
}\\
\subfigure{
\begin{minipage}[t]{0.4\textwidth}
\includegraphics[width=1\linewidth]{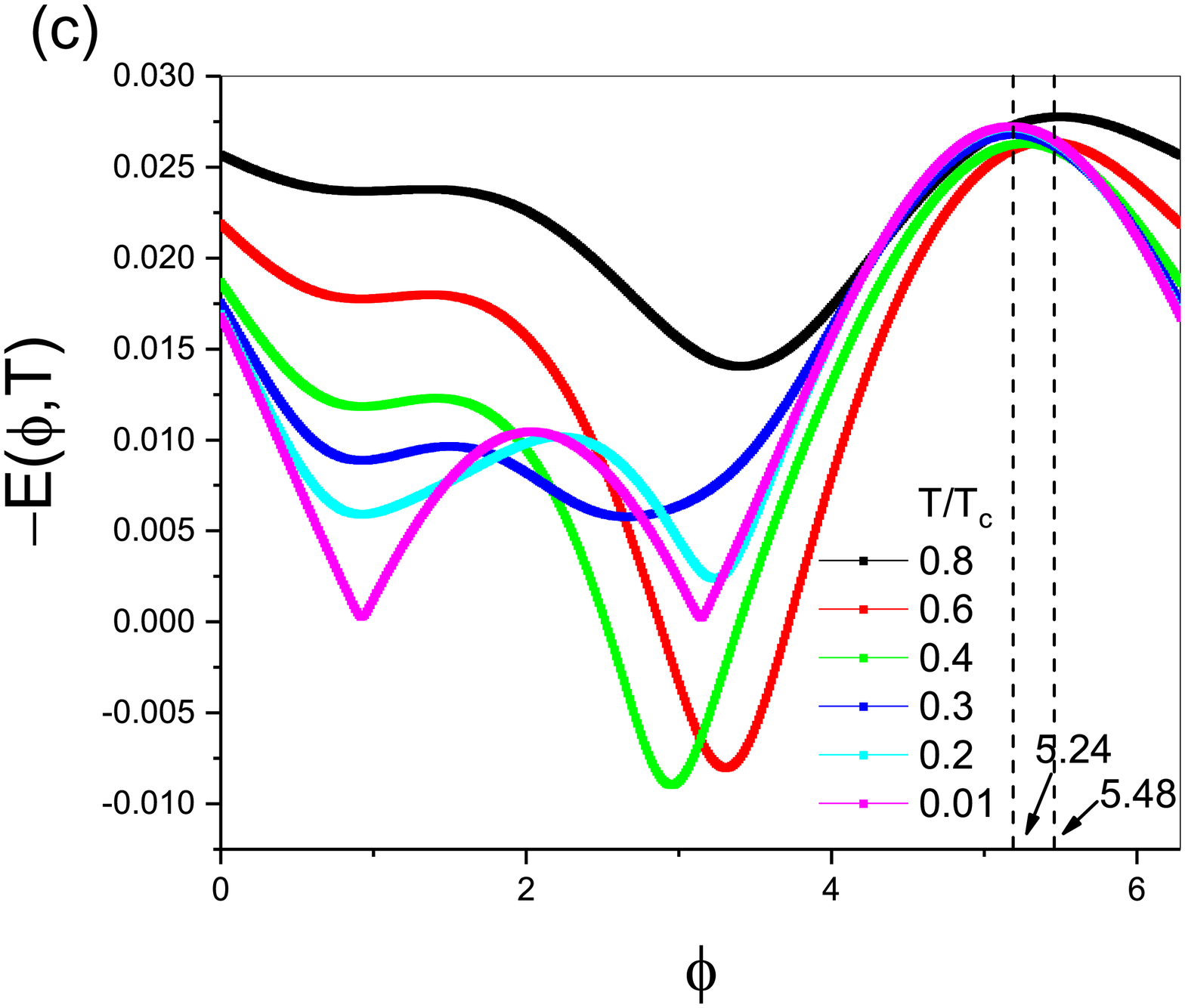}
\label{fig:side:a}
\end{minipage}
}
\caption{(Color online) Free energy obtained by the approximated Andreev level in Eq.(16).
where we show the free energy in one period.
The temperature is setted as 1.5 $T_{c}$.
The corresponding setting of the parameters are labeled in the plot.
(b)-(c) show the free energy with different temperatures,
while the other parameters as setted as:
$E_{\perp}=0.034$ eV, $\mathcal{A}=0.3$, $\mu_{n}=0.2$ eV, $L=50$ \AA.
The dash-line in (b)-(c) indicate the position of minimum free energy.
}
\end{figure}
\clearpage
Fig.8
\begin{figure}[!ht]
   \centering
   \centering
   \begin{center}
     \includegraphics*[width=0.5\linewidth]{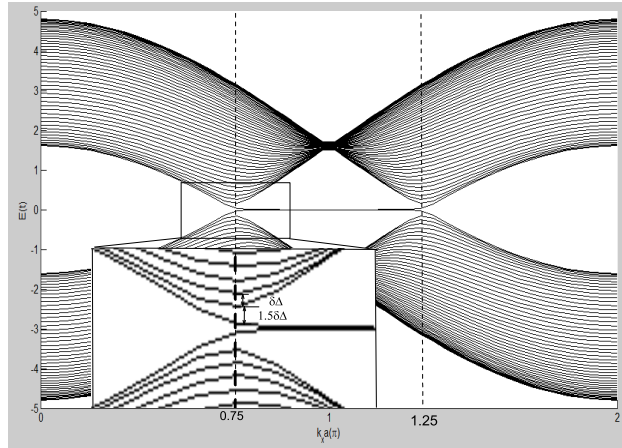}
\caption{Band structure of the zigzag silicene nanoribbon in
a strip geometry. The energy $E$ is in unit of $t$. 
The middle band at $E=0$ is the zero-model edge (Dirac level) which localed in the range $[0.75\pi,1.25\pi]$.
The zero-model level is at $E=0$ for zero chemical potential and zero electrostatic potential which guarantees the electron-hole symmetry.
Inset: the enlarged view near the valley K at $k_{x}a=0.75\pi$.
}
   \end{center}
\end{figure}

\end{document}